\documentclass[aps,prl,twocolumn,showpacs,superscriptaddress]{revtex4}
\usepackage{graphicx}

\usepackage{dcolumn}
\usepackage{bm}
\begin{document}

\title{Nanowire Acting as a Superconducting Quantum Interference Device}

\author{A.~Johansson}
\affiliation{Department of Condensed Matter Physics, The Weizmann Institute of Science, Rehovot 76100, Israel}
\author{G.~Sambandamurthy} 
\thanks{Present address: National High Magnetic Field Laboratory, Tallahassee, FL 32306, USA}
\affiliation{Department of Condensed Matter Physics, The Weizmann Institute of Science, Rehovot 76100, Israel}
\author{N.~Jacobson}
\affiliation{Department of Materials and Interfaces, The Weizmann Institute of Science, Rehovot 76100, Israel}
\author{D.~Shahar}
\affiliation{Department of Condensed Matter Physics, The Weizmann Institute of Science, Rehovot 76100, Israel}
\author{ R.~Tenne}
\affiliation{Department of Materials and Interfaces, The Weizmann Institute of Science, Rehovot 76100, Israel}
\date{\today}

\begin{abstract}
We present the results from an experimental study of the magneto-transport of superconducting wires of amorphous Indium-Oxide, having widths in the range 40 -- 120 nm. We find that, below the superconducting transition temperature, the wires exhibit clear, reproducible, oscillations in their resistance as a function of magnetic field. The oscillations are reminiscent of those which underlie the operation of a superconducting quantum interference device.
\end{abstract}


\maketitle

 
The central challenge in the study of thin superconducting wires is to understand how superconductivity is affected when approaching the one-dimensional (1D) limit. Earlier studies have predicted that intrinsic thermal \cite{Langer1967,McCumber1970} and quantum \cite{Giordano1988,Zaikin1997,Golubev2001} fluctuations play an increasingly important role in this limit, causing the wires to remain resistive much below the superconducting transition temperature, $T_{c}$. Recent theories \cite{Oreg1999, Smith2001}, incorporating the effect of electron-electron interactions, describe the suppression of $T_{c}$ when approaching the 1D limit, which was observed in experiments \cite{Graybeal1987}. 

In recent years new experimental techniques enabling the fabrication of superconducting wires with a diameter approaching the 1D limit were developed. The experiments that followed \cite{Bezryadin2000}
focused primarily on whether the quantum resistance for a Cooper-pair, $R_{Q} = h/4e^{2} \approx 6.45$ k$\Omega$ ($h$ is Planck's constant and $e$ is the charge of the electron) is the resistance scale that solely controls the existence of superconductivity. While some works \cite{Bezryadin2000, Bollinger2004} provided evidence that wires with a normal state resistance $R_{N}<R_{Q}$  are superconducting and those with $R_{N}>R_{Q}$ become insulating at low $T$, this point is still under debate \cite{Lau2001}. 

In this letter we report on an experimental study of the magnetic field ($B$) dependence of the resistance of superconducting wires whose dimensions are close to the 1D limit. We find that, while a strong $B$ drives our wires into an insulating state, the magnetoresistance is dominated by reproducible periodic oscillations similar to those observed in superconducting quantum interference devices (SQUIDs) \cite{Tesche1977}. We also find that wires with $R_{N}>>R_{Q}$ can exhibit superconductivity.

In order to fabricate our 1D wires we utilized the method of Bezryadin {\it et al.} \cite{Bezryadin2000}, in which a non-conducting nanotube, suspended across a narrow gap etched in a semiconductor substrate, is used as a template on which the superconductor is deposited. 
There are two experimental differences between our work and that of Ref. \cite{Bezryadin2000}. First, instead of carbon, our nanotubes were made of WS$_{2}$ \cite{Tenne1992}. Being a semiconductor with a bandgap of about 2 eV \cite{Frey1998}, WS$_{2}$ nanotubes are electrically insulating at low $T$ and do not create a parallel conduction channel. We have verified that the nanotubes are insulating before depositing the superconducting material. 

Second, and more importantly, for the disordered superconductor we chose amorphous indium-oxide (a:InO). 
This choice was influenced by several of its properties. Since a:InO is known to form relatively uniform, superconducting films \cite{Hebard1982}, we expected wires made from it to be uniform as well. 
Also, a:InO was used extensively to study superconductivity in two-dimensions (2D) (see Ref. \cite{Sambandamurthy2004} and references therein), including the interplay between disorder, superconductivity and an external magnetic field. Finally, by a simple process of thermal annealing, the level of disorder can be changed continuously \cite{Fiory1983} allowing fine-tuning of the samples to the desired disorder level for the low-$T$ experiments. 

The inset of Fig. 1 depicts a typical device. It is built on an intrinsic GaAs wafer with electrodes of $Au$ evaporated on the surface. A 4 $\mu m$ deep and 1 $\mu m$ wide trench is etched between the electrodes across which a 
WS$_{2}$ nanotube, from an isopropanol suspension, is trapped using dielectrophoresis. Thereafter a:InO is e-gun evaporated in a high vacuum ($10^{-7}$ Torr) system. The deep trench between the electrodes prevents the a:InO layer from shorting the electrodes. The a:InO wire has the same width and length as the part of the nanotube streching across the trench. Transmission electron microscopy revealed that the InO evaporated on top of 
WS$_{2}$ nanotubes is amorphous, similar to when it is prepared as films.
The devices were measured at low $T$s in either a He3 refrigerator with a base $T$ of 0.23 K, or a dilution refrigerator attaining 0.01 K. 
Resistance measurements were performed using standard low-frequency (4 - 14 Hz) lock-in techniques employing home-built low-noise voltage preamplifiers. Low excitation currents in the range of 0.1 - 1 nA were used to minimize heating of the wires. 

\begin{figure}
\includegraphics[width=8cm]{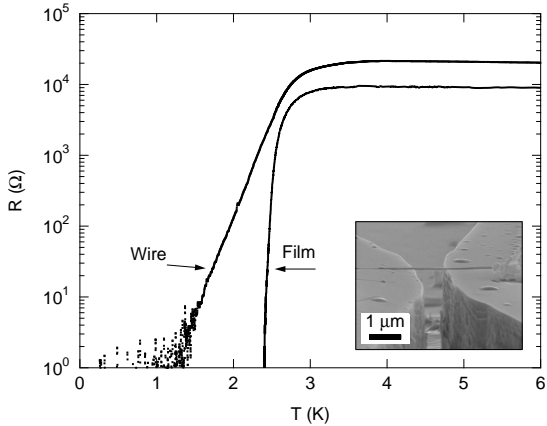}
\caption{\label{}$R$ vs. $T$ obtained from wire 99Fe whose width is 100 nm, together with data from a similarly prepared 500 $\mu$m-wide film. $T_{c}^{wire}=2.8$ K and $T_{c}^{film} = 2.7$ K. Inset: An SEM image of a typical device.}
\end{figure}

In Fig. 1 we plot $R$ vs. $T$ obtained from one of our wires. The general aspects of these data are typical of most of our superconducting wires.  In the vicinity of $T=4$ K, $R$ assumes a (shallow) maximum value, which will be referred to as the normal-state resistance, $R_{N} =21.4$ k$\Omega$ for this wire. 
We note, see Table \ref{Wire parameters}, that $R_{N}$ of our superconducting wires can be as high as $131$ k$\Omega$, which is much higher than observed in previous studies of superconducting wires \cite{Giordano1990,Sharifi1993,Bezryadin2000,Lau2001}.

As $T$ is lowered below 4 K, $R$ begins a rapid decrease, indicating that the wire is entering a superconducting state. The critical temperature of superconductivity, defined by the temperature at which $R$ is reduced to half of $R_{N}$, 
is here $T_{c}=2.8$ K. The transition is broad, $\Delta T= 0.9$ K, 
significantly more so than in similarly prepared 2D films, signaling the approach to the 1D limit \cite{Giordano1990,Sharifi1993}. 
In some of our wires, the resistance ceases to drop and saturates at a 
low-$T$ value between 0.06-40 k$\Omega$. Other wires, such as the wire in Fig. 1, show a vanishing resistance in the low-$T$ limit. While the presence of residual resistance far below $T_{c}$ could be a signature of resistive phase-slip processes due to quantum fluctuations, \cite{Giordano1988,Zaikin1997,Golubev2001,Bezryadin2000,Lau2001} more mundane reasons related to a possible decoupling of the electrons from the thermal bath are also possible. We intend to address this issue in another study. 

\begin{figure}
\includegraphics[width=8cm]{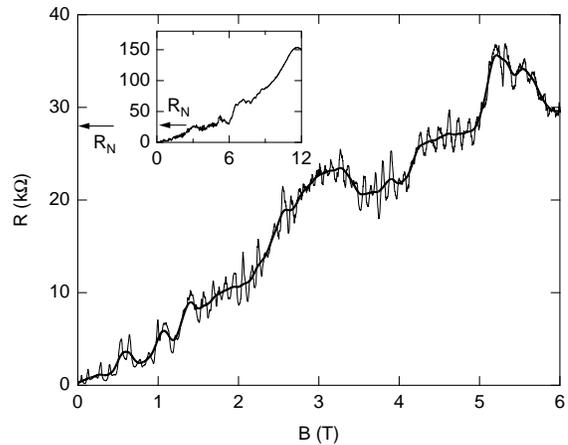}
\caption{\label{}$R$ vs. $B$ at $T = 0.012$ K for wire 99Nb. A smooth background, shown by the bold curve, is subtracted from the data to obtain the oscillatory part shown in Fig. 3(a). Inset: $R$ versus $B$ displayed up to our highest accessible $B$-field of 12 T.}
\end{figure}

For the rest of this Letter, we focus on the $B$-dependence \cite{Rogachev2005} of the resistance of our wires.
In the inset of Fig. 2 we show $R$ vs. $B$ obtained from wire 99Nb, with the $B$-field applied perpendicular to the wire and the contact pads. Typical of all our superconducting wires, $R$ increases with $B$ reaching a maximum value of $154$ k$\Omega$ at $B=11.6$ T. This value is significantly larger than $R_{N}$ = 28.4 k$\Omega$ indicating that, in addition to the complete destruction of superconductivity, the high $B$ has caused the wire to become insulating. 

The central result of this work is shown in the main part of Fig. 2, where we plot the low-$B$ range of the data presented in the inset of Fig. 2. Inspecting this figure one can see that, superimposed on top of the resistance increase as a function of $B$, are clear and well-resolved resistance oscillations. The oscillations are reproducible, and constitute an intriguing feature in a superconductor with an apparent singly connected geometry. To extract the oscillatory part of $R$ versus $B$, $R_{osc}$, we subtract a smooth background from $R$, shown by the bold curve in Fig. 2. The result is displayed in the top trace of Fig. 4(a). Although largely fluctuating, the amplitude of the oscillations does not show a systematic dependence on $B$ up to $\approx 6$ T, above which significant oscillations are no longer detected. In the $B$ range from 0 -- 6 T the average rms value of the oscillations is $0.9 \pm 0.2$ k$\Omega$. When performing a fast fourier transform on the data in Fig. 4(a), the regular oscillations manifest a prominent peak in the interval of 8.3 -- 9.8 1/T (see Fig. 4(c)).

\begin{figure}
\includegraphics[width=8cm]{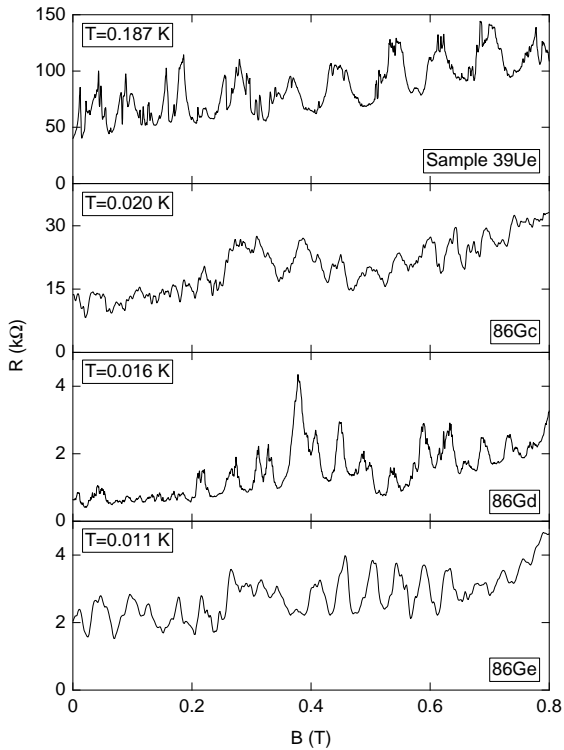}
\caption{\label{}$R$ vs. $B$ data from four states, displayed in a narrow $B$-range for clarity. }
\end{figure}

We have found similar oscillations in most of our superconducting wires. In Fig. 3, we plot the $R$ versus $B$ applied perpendicular to the wires for samples 39Ue, 86Gc, 86Gd and 86Ge, of which the latter three are different annealing stages of the same physical wire. 
Although the samples vary widely in their $R_{N}$, residual resistance and rms-values of their oscillations (note the different scales of resistance in the figure), the oscillation periods are surprisingly close to each other, all in the range 0.043 -- 0.083 T.

\begin{figure}
\includegraphics[width=8cm]{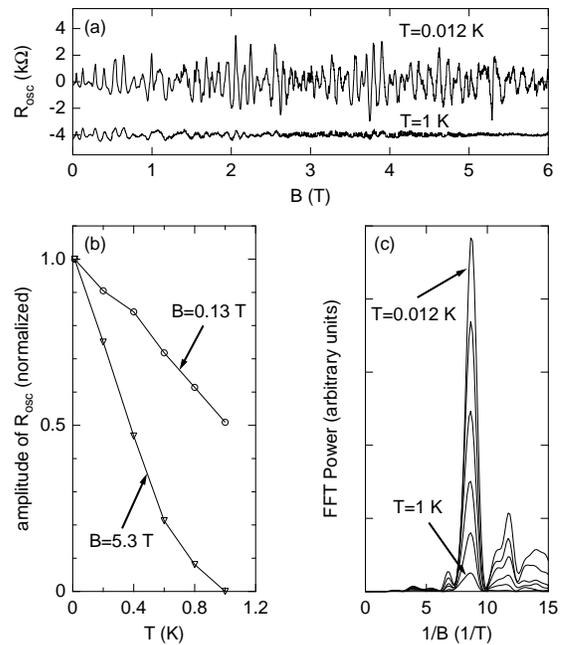}
\caption{\label{}(a) $R_{osc}$ vs. $B$ taken at $T=0.012$ (top trace) and 1 K (shifted for clarity by $-4$ k$\Omega$), from wire 99Nb. (b) $T$-dependence of the amplitude of $R_{osc}$ vs. $B$ peaks at $B=0.13$ T and $B=5.3$ T, normalized to their value at $T=$0.012 K. (c) FFT traces obtained from isotherms of $R_{osc}$ vs. $B$ in the interval 0 -- 1.7 T, taken at $T=$0.012, 0.2, 0.4, 0.6, 0.8, and 1 K. The location of the dominant peak, 8.6 1/T, corresponding to the main frequency of the oscillations, remains constant over the entire $T$-range, while its height decreases sharply.} 
\end{figure}

To study the $T$-dependence of the oscillations we measured $R$ vs. $B$ isotherms of wire 99Nb at several $T$'s. In Fig. 4(a) we plot $R_{osc}$ obtained at $T=0.012$ and 1K. We find that the amplitude of the oscillations is suppressed with increasing $T$. The suppression is $B$-dependent and is weaker at lower $B$ where the oscillations are still observed even at $T=1$ K. To demonstrate this, we plot in Fig. 4(b) the relative suppression of the  amplitude of $R_{osc}$ with increasing $T$ for the two peaks situated at $B=0.13$ T and $B=5.3$ T. The peak at low $B$ is still visible at $T=$1 K with an amplitude close to half its amplitude at $T=$0.012 K, while the high $B$ peak is fully suppressed.

Another significant feature of the resistance oscillations can be seen in Fig. 4(c), where we plot FFT traces obtained from data in the $B$ range from 0 to 1.7 T. Although the amplitude of the dominant peak is strongly $T$-dependent, its position is not, showing that the period of the oscillations is independent of $T$. If one period of the oscillations corresponds to a quantum of flux threading a coherent area of the sample, the $T$-independent period may indicate that, rather than an intrinsic length scale, a geometrical factor is setting the relevant area for the oscillations.  

\begin{table*}
\caption{\label{Wire parameters} Parameters for wires 39Ue, 86Gc, 86Gd, 86Ge, and 99Nb, all showing oscillatory behavior in $R$ vs. $B$. $t$ is thickness of the a:InO deposition. $w$ and $L$ are width and length of wires estimated from SEM images, and have an uncertainty of $\pm$10 \%. $R_{N}$ is the normal state resistance. $R_{res}$, the residual resistance, and rms-value of $R_{osc}$ were measured at $T=$0.187, 0.020, 0.016, 0.011, and 0.012 K, respectively. $f_{osc}$ is the dominant frequency for oscillations found from FFT of $R$ vs $B$ data. All wires have a $T_{c}$ in the $T$-interval 2.5 -- 3 K.}
\begin{ruledtabular}
\begin{tabular}{|c|c|c|c|c|c|c|c|}
wire & $t$ (nm) & $w$ (nm) & $L$ ($\mu m$) & $R_{N}$ (k$\Omega$) & $R_{res}$ (k$\Omega$) & rms $R_{osc}$ (k$\Omega$) & $f_{osc}$ (1/T) \\ \hline \hline
39Ue & 21 & 40 & 1.5 & 55.1 & 36.9 & 15.5 & 12.1 \\  \hline
86Gc & 30 & 94 & 1.0 & 131 & 9.51 & 2.30 & 20.7 \\  \hline
86Gd & 30 & 94 & 1.0 & 122 & 0.30 & 0.36 & 21.6 \\  \hline
86Ge & 30 & 94 & 1.0 & 115 & 1.31 & 0.46 & 23.0 \\  \hline
99Nb & 30 & 120 & 3.4 & 28.4 & 0.04 & 0.90 & 8.6 \\  
\end{tabular}
\end{ruledtabular}
\end{table*}

This situation is reminiscent of an earlier study, by Herzog {\it et al.} \cite{Herzog1998}, who reported magnetoresistance oscillations in granular superconducting wires of Sn, where the extracted normal area, $A_{N}$, was close to the average grain size in their wires. The oscillations were attributed to the effects of screening currents around phase-coherent loops of weakly linked superconducting grains. Our TEM studies show no indication of granularity in our wire. Moreover, our extracted $A_{N}$s are about an order of magnitude larger, and are comparable to the area of our wires, $A_{\bot}$. For most of our wires $A_{N}/A_{\bot} \approx 0.5$ alluding to an orbital mechanism that is coherent over the entire length of the wires. Only in wire 99Nb, which is the longest wire we studied, the oscillations are not periodic in the flux through half of $A_{\bot}$, and exhibit a slower periodicity, showing that coherence is no longer maintained over the whole wire.

The appearance of resistance oscillations as a function of $B$ in a superconductor is not necessarily surprising. Such oscillations are the basis of the operation of SQUIDs. In essence, the oscillatory dependence on $B$ is enabled because the annular geometry of the SQUID allows for a place for the superconducting vortices to reside with minimal cost in energy. In our wires no such geometry exists and the origin of the SQUID-like oscillations in our data is not fully understood. It is possible that vortex cores energetically prefer to locate at the center of the wire, and supercurrents can still flow unhindered near the edge, rendering the central region normal and inducing an effective multiply-connected geometry. 

To summarize, we have reported on oscillatons in the magnetoresistance of thin superconducting wires with an apparent singly connected geometry.  The oscillations are reminiscent of those of a SQUID. The amplitude of the oscillations is strongly $T$ dependent and is almost fully suppressed at $T$=1 K, while the frequency shows no $T$ dependence. We find the oscillations in $R$ versus $B$ data from several wires, which span a wide range in residual resistance at low $T$'s. 

We thank Yuval Oreg and Kathryn A. Moler for illuminating discussions, Rita Rosentsveig for synthesizing the WS$_{2}$ nanotubes and Ronit Popovitz-Biro for help with characterization of the a:InO deposition. This work is supported by the ISF, the Koshland Fund and the Minerva Foundation.


\end{document}